\documentclass[manuscript]{aastex}
\usepackage{spr-astr-addons}
\usepackage{url}
\usepackage[utf8]{inputenc}
\usepackage[T1]{fontenc}

\RequirePackage{color}

\newcommand{\emaila}{jameel@giki.edu.pk}

\begin{document}

\title{Mass fractions in stellar interior during presupernova evolution}

\author{Jameel-Un Nabi\altaffilmark{1}}
\affil{Faculty of Engineering Sciences, GIK Institute of Engineering
Sciences and Technology, Topi 23640, Swabi, Khyber Pakhtunkhwa,
Pakistan} \email{\emaila}
\author{Abdel Nasser Tawfik\altaffilmark{2}}
\affil{Egyptian Center for Theoretical Physics, MTI Modern
University, Al Mokattam, 11571 Cairo, Egypt}

\author{Nada Ezzelarab\altaffilmark{2}} \affil{Egyptian Center for Theoretical Physics, MTI Modern
University, Al Mokattam, 11571 Cairo, Egypt} \and
\author{Ali Abas Khan\altaffilmark{1}}
\affil{GIK Institute of Engineering Sciences and Technology, Topi
23640, Khyber Pakhtunkhwa, Pakistan} \altaffiltext{1}{Corresponding
author email : jameel@giki.edu.pk}

\sloppy
\begin{abstract}
We assume nuclear statistical equilibrium (NSE) conditions and use
Saha Equation to calculate mass fractions in stellar interior during
presupernova evolution of massive stars. Our ensemble contains 728
nuclei. The distinguishing feature of our calculation is a
state-by-state summation of nuclear level densities up to 10 MeV for
all nuclei considered in our ensemble. This leads to a more
realistic description of nuclear partition function which poses one
of the largest uncertainties in the mass fraction calculations under
conditions of NSE. The Coulomb correction term was taken into
account for the calculation of ground-state binding energies of the
nucleons. Our calculated mass fractions are up to four orders of
magnitude smaller than previous calculation. The current calculation
could prove useful for study of stellar matter during presupernova
phases of massive stars.
\end{abstract}

\keywords{Nuclear Statistical Equilibrium, Nuclear Level Density,
Nuclear Partition Function, Nuclear Abundance, Mass Fractions,
Stellar Core-Collapse.}

\section{Introduction}

The study of various models of the explosive nucleosynthesis as well
as the $r$-process \cite{Ptitsyn,Woosley,Nadyozhin} requires
knowledge of the chemical composition close to the nuclear
statistical equilibrium (NSE), which is as an initial condition for
nuclear reaction network calculations. In NSE scenario, an
equilibrium is achieved between strong and electro-magnetic
interactions for all sort of nuclear species. The weak interaction,
however, is not in equilibrium as the stellar matter is transparent
to neutrinos produced and energy is channeled out of the system.
Under the assumption of NSE, the isotropic abundances of nuclei
follow simply from nuclear Saha's equation for a given density,
temperature, and electron fraction \cite{Hartmann}.  Under NSE
condition, each nucleus can be transformable to any other nucleus
and the radiation and matter must be in equilibrium in the stellar
assembly. Details of NSE conditions can be seen in \cite{Hoyle}.
During stellar evolution, when silicon burning terminates, all
nuclear reactions are studied by the NSE. A number of authors
studied the problem of NSE by using Saha's type equations
\cite{Hartmann, Liu, Clifford, Epstein, Aufderheide, Bravo}.

The thermodynamic conditions spanned over several $100$ ms time
interval post bounce in core-collapse supernova simulation have been
analyzed \cite{Fis14}, where the density range spanned about $10$
orders of magnitude (10$^{5}$ - 10$^{15}$ g cm$^{-3}$), the
temperature around two orders (0.1 - 50 MeV) and the lepton fraction
more than an order of magnitude (0.01 - 0.5). The large variety of
nuclear matter properties leads to distinct outcomes in supernova
simulations. The rate of change of lepton-to-baryon fraction ($Y_e$)
depends both on mass fractions and weak-interaction rates of the
nuclei. The acceleration of core-collapse of massive stars changes
with NSE. Recently we studied the evolution of composition of
nuclear matter over a wide temperature (10$^{9}$ -- 10$^{11}$ K) and
density range (10$^{1}$ - 10$^{11}$ g cm$^{-3}$) \cite{Nab15}. In
the present paper we study the mass fractions of nuclei and free
nucleons as a function of $Y_e$ of the stellar matter. Our ensemble
contained a total of 728 nuclei, having mass number up to 100 which
were selected primarily to cover a decent number of nuclei for
astrophysical simulations.

Accurate and reliable values of nuclear partition function (NPF) are
a prerequisite for the determination of the mass fractions in the
presupernova matter. NPF also plays key role in other astrophysical
processes, e.g. determination of equation of state during
gravitational collapse \cite{Ful92}  and abundances isotopic
calculations in silicon burning \cite{Hix96}. In a recent study of
isotopic abundances in presupernova cores \cite{Dim02}, it was
concluded that low-lying nuclear states need to be treated
preferably as discrete and a summation over these states lead to
significant change in the calculated value of NPF which in turn
poses one of the biggest uncertainty in the calculation of mass
fractions. The calculated NPF by the authors showed a deviation of
up to 50$\%$ when discrete energy levels were considered as against
those assuming a level density function and performing integrals to
obtain the NPF. However the authors in \cite{Dim02} were able to
perform summation over discrete energy levels only up to 3 MeV. We
performed summation for all discrete energy levels up to 10 MeV in
our present calculation. The pn-QRPA model was used to calculate all
discrete energy levels up to 10 MeV. Soaring stellar temperatures
prior to collapse gives a finite contribution to NPF even up to 10
MeV. Beyond 10 MeV we do assume a level density function in our
calculation of NPF. Further to this, authors in \cite{Dim02} used a
total of 65 nuclear species in their calculation whereas the current
paper deals with a much bigger ensemble of 728 nuclei.

The present paper is organized as follows. Section 2 briefly
describes the formalism which we adopted to calculate mass
fractions, nuclear level density and NPF. Section 3 discusses our
results and also compares our calculation with previous calculation.
We finally conclude our findings in Section 4.

\section{Formalism}
Our NSE treatment is similar to that employed by \cite{Hartmann,
Clifford, Aufderheide, Kod75}. Under NSE the mass fraction for
nuclei other than neutrons and protons is given by the Saha equation
\small{
\begin{eqnarray}\label{x1}
X(A,Z)= \frac {G(A,Z,T)}{2}\left(\frac{\rho_{st} N_{0}
\lambda_{T}^{3}}{2}\right)^{A-1}\times \\ \nonumber
A^{\frac{5}{2}}{X_{n}^{A-Z}X_{p}^{Z}}\exp \left[ B(A,Z)/kT \right].
\end{eqnarray}
\normalsize
where $X_{n}$ and $X_{p}$ are the mass fractions of free neutron and
protons, respectively, $G(A,Z,T)$ is the temperature-dependent
nuclear partition function, $B(A,Z)$ is the ground state binding
energy, $T$ is stellar temperature, $\lambda_{T}$ is the thermal
wavelength, $k$ is the Boltzmann constant, $A$ is the mass number,
$\rho_{st}$ is the stellar density and $N_{0}$ is Avogadro's number.
Neglecting the Coulomb's correction, the ground state binding energy
$B(A,Z)$ is given by
\begin{equation}\label{be}
B(A,Z) = c^{2}[N m_{n} + Z m_{p} - m (A,Z)],
\end{equation}
where $m_{n}$, $m_{p}$ and $m(A,Z)$ denote the masses of free
neutrons, protons and nuclei, respectively. Due to electron
screening the nuclear binding energies in stellar environment should
be different than those in the vacuum \cite{Ish03}. Consequently,
driplines are shifted and the heavy nuclei become unstable against
fission under terrestrial condition. However they may become stable
in supernova matter. These might be very important under
astrophysical conditions. We included the Coulomb's correction in
binding energy equation using Hartree-Fock approximation in
Wigner-Seitz cells \cite{Chamel}.  When Coulomb's correction is take
into account the ground state binding energy gets the form
\begin{equation}\label{bf}
B(A,Z)^{cc} = B(A,Z)-\triangle{V^{cc}_{f}(\rho_{e})},
\end{equation}
where
\begin{equation}\label{bl}
\triangle{V^{cc}_{f}(\rho_{e})} = -\frac{3}{5}\frac
{Z^{2}e^{2}}{R_{0}}\bigg(\frac{3}{2}\eta-\frac{1}{2}\eta^{3}\bigg),
\end{equation}
here $\eta = \frac{n_{e}}{n_{0}}\frac{A}{Z}$ and $n_{e}$ represents
free electron density. We assume nucleons as a homogeneous sphere of
radius $R_{0}$ at a maximum density $n_{0}$. Radius of the sphere is
given by
\begin{equation}
R_{0} = \bigg(\frac{3A}{4\pi n_{0}}\bigg)^{1/3}.
\end{equation}
Temperature dependent NPF plays a key role in calculation of
abundance of nuclear species.   An improvement was made in this
study to use a true temperature dependent NPF, $G(A,Z,T)$, compared
to previous calculations of NPF e.g. \cite{Epstein}. The NPF over
discrete energy levels can be calculated using
\begin{equation}\label{p1}
G(A,Z,T)= \sum_{i}(2J_{i}+1) \exp[{-E_{i}/kT}],
\end{equation}
here $J_{i}$ and $E_{i}$ denotes spin and energy of the $i^{th}$
state. However it is a well-known fact that the discrete energy
levels become continuous as the excitation energy increases and then
a more generalized nuclear partition function with level densities
normalized to ground state energy is given by \cite{Fow67}:
\small{
\begin{align}\label{ld}
G(A,Z,T)&=\sum_{\mu =0}^{\mu_{m}}(2J^{\mu}+1)\exp[{-E^{\mu}/kT}]+\nonumber\\
&\int_{E^{\mu_{m}}}^{E^{max}}\sum_{J^{\mu},\pi^{\mu}}(2J^{\mu}+1)\exp(-\epsilon/kT)\rho(\epsilon,J^{\mu},\pi^{\mu})d\epsilon,
\end{align}}
\normalsize
where $\rho$($\epsilon$, $J^{\mu}$,$\pi^{\mu})$ denotes the level
density function and $\mu_{m}$ is the label of the last employed
either experimentally known or theoretically estimated energy state.
The sum runs over all thermally discrete excited states up to
$\mu_{m}$. In Eq. (\ref{ld}) the first term shows contribution from
experimentally known low energy states  as well as pn-QRPA
calculated theoretical energy eigenvalues up to $E^{\mu}$ while
second term denotes contribution from remaining continuous states.
The continuous excited states contribution was calculated using
level density formula $\rho$($\epsilon$, $J^{\mu}$,$\pi^{\mu})$.  We
used the pn-QRPA model to calculate all discrete energy levels, up
to 10 MeV, for all nuclide under consideration. Formalism for
calculation of nuclear energy levels within the pn-QRPA formalism
may be seen from \newline \cite{Mut92, Nab99, Nab04} and is not reproduced
here for space consideration. We incorporated the latest
experimental data into our calculation for increased reliability,
wherever possible. The calculated excitation energy was replaced
with an experimental one when they were within $0.5$ MeV of each
other (missing measured states were inserted) along with their
$2J+1$ values. For a particular nucleus, for measured (inserted)
states with the missing angular momentum, the highest reported value
of $2J+1$ was used. For cases where more than one value of $J$ was
reported, we selected the highest value.

Beyond 10 MeV energy scale and by assuming a uniform Fermi gas we
estimated the level density (using saddle point approximation yields
and some additional simplifications) as \cite{Boh69}
\begin{equation}\label{lda}
\rho(A,E) = \frac{\exp(2\sqrt{aE})}{4\sqrt{3}E},
\end{equation}
where $a = \frac{\pi^{2}g}{6}$ and the density of single particle
states for a nucleus having $A$ fermions is $g =
\frac{3A}{2\epsilon_{f}}$ and $\epsilon_{f}$ is the Fermi energy. Of
course a more sophisticated treatment for calculation of the nuclear
level density function is possible (e.g. back-shifted, pairing
corrections, inclusion of spin cut-off parameters). It is to be
noted that we microscopically calculated all energy levels up to
$10$ MeV in daughter. Beyond $10$ MeV, the approximate formula used
in Eq.~(\ref{lda}) still gives us a reliable estimate of the level
density for calculation of temperature-dependent nuclear partition
functions  to finally calculate the mass fractions in stellar matter
(Eq.~(\ref{x1})).

The thermally distributed nuclei must conserve stellar mass
\begin{equation}\label{mass}
\sum_{k}{X_{k}(A,Z)}=1.
\end{equation}
Further conservation of charge demands
\begin{equation}
\sum_{k}\frac{X_{k}Z_{k}}{A_{k}}=\frac{(1-\eta)}{2}= Y_{e},
\end{equation}
where $Y_{e}$ depends on the electron number density given by
$n_{e}=N_{0}\rho Y_{e}$.

\section{Results and Comparison}
Fig.~\ref{fig0} depicts the nuclei present in our ensemble. There
were a total of 728 $sd$-, $fp$ and $fpg$-shell nuclei present in
our pool for calculation of mass fractions using the Saha Equation
(Eq.~(\ref{x1})). The list of nuclei was inspired by the earlier
calculation of weak-interaction rates in stellar matter by Nabi and
Klapdor-Kleingrothaus \cite{Nab99, Nab04}.

The nuclear level density (NLD) is a key ingredient in the
calculation of nuclear reaction rates in astrophysics. Typical
outcomes of NLD calculation for nuclei in the Fe-region are shown in
Fig.~\ref{fig1} and Fig.\ref{fig2}. These seven nuclei are the same
as the seven most abundant nuclei in NSE calculated by \cite{Liu} at
$T = 5\times10^{9}$K and $\rho_{st} = 1\times10^{7-9}$ g/cm$^{3}$.
It can be seen from the figures that $^{58}$Ni has the maximum level
density and $^{52}$Cr the least which is also in agreement with the
assumed level density function Eq.~(\ref{lda}).

The temperature-dependent NPFs for few iron-regime nuclei as
function of the stellar temperature $T_{9}$ (which is measured in
units of $10^{9}$K) are shown in Table~\ref{ta1}. Our calculation is
also compared with the statistical model calculation of
\cite{Rau00}. The Hauser-Feshbach formalism was used by \cite{Rau00}
to perform a statistical model calculation of level densities and
associated nuclear partition functions. The finite range droplet
model (FRDM) \cite{Moe95} was used as the input mass model to
calculate the temperature-dependent NPFs by \cite{Rau00}.   Whereas
the two NPF calculations compare well for even-even nuclei,
discrepancies of one order of magnitude or more are seen for odd-A
and odd-odd cases. Authors in \cite{Rau00} applied the statistical
description throughout the nuclear chart without relying on
experimental level density parameters in specific cases. The authors
did comment that this feature of their formalism may lead to locally
larger deviations from experiment. The microscopic calculation of
theoretical energy levels (up to 10 MeV) and insertion of
experimental energy levels resulted in a reliable and accurate
estimate of NPF in our calculation. It is noted that the value of
NPF increases as stellar temperature soars to high values. The rate
of change of NPF increases with increasing temperatures. These
effects may be traced down to Eq.~(\ref{ld}).

We present our calculation of mass fractions in Fig.~\ref{fig3} and
Fig.~\ref{fig4}. The calculated mass fractions of eight most
abundant nuclei at $T = 5\times10^{9}$K and $\rho_{st} =
1\times10^{7}$ g/cm$^{3}$ as a function of $Y_{e}$ is shown in
Fig.~\ref{fig3} and may be contrasted with Fig.~1 of \cite{Liu}. It
is seen from our Fig.~\ref{fig3} that the mass fractions increases
monotonically with lepton content of the stellar matter. The
difference with the calculation of \cite{Liu} is attributed to a
more rigorous treatment of discrete energy levels up to 10 MeV for
calculation of NPF, consideration of Coulomb correction in the
binding energy of the nucleons and the choice of a much larger
ensemble of nuclei in our mass fraction calculation. Fig.~\ref{fig4}
depicts the evolution of mass fraction calculation of our eight most
abundant nuclei at $T = 5\times10^{9}$K and a higher stellar density
of $\rho_{st} = 1\times10^{9}$ g/cm$^{3}$ with increasing $Y_{e}$.
Our Fig.~\ref{fig4} may be compared with Fig.~2 of \cite{Liu}. It is
noted in Fig.~\ref{fig4} that there is an inclusion of heavy
nucleus, $^{92}$Mo, in the list of most abundant nuclei as stellar
core stiffens to higher density. Moreover alphas are also very
abundant at high stellar densities according to our calculation. A
direct comparison of our calculation with those of \cite{Liu} can be
seen in Table~\ref{ta2}. The comparison of calculated mass fractions
is shown at $T = 5\times10^{9}$K and $\rho_{st} = 1\times10^{7}$
g/cm$^{3}$ and $Y_{e} = 0.5$. It is evident from Table~\ref{ta2}
that our model with new features leads to a much smaller  calculated
value of mass fractions (up to four orders of magnitude). In order
to understand the origin of differences in the two calculations we
lift some of the features of our calculation in order to come close
to the conditions adopted by \cite{Liu} in their calculation of mass
fraction. This new set of calculation can be seen in column~3 of
Table~\ref{ta2}. In order to perform our calculation with revised
features we (i) summed the energy levels up to an excitation energy
of 3 MeV, (ii) did not employ the Coulomb correction term in
calculation of binding energies and (iii) chose a much smaller
ensemble containing only iron-group nuclei in mass range 50 - 60.
The smaller pool used for calculation of mass fractions containing
only 89 nuclei can be seen in Fig.~\ref{fig5}. We note that the two
calculations still show differences. In order to further understand
the origin of these differences we compare first the calculation of
temperature-dependent NPFs used in the two mass fraction
calculations for these six nuclei. Authors in \cite{Liu} used the
calculated NPFs of \cite{Rau00} till $T_{9}$ = 10. For still higher
stellar temperatures they employed the NPF of \cite{Rau03}. We
compare the temperature-dependent NPFs of the six nuclei with those
of \cite{Rau00} in Table~\ref{ta3} and with those of \cite{Rau03} in
Table~\ref{ta4}.  In addition to FRDM model, the extended
Thomas-Fermi approach with Strutinski Integral (ETFSI) \cite{Pea96}
model was used by Rauscher as input mass formula for calculation of
his NPFs at high temperatures. In Table~\ref{ta4} we compare our
results with both mass models employed by Rauscher. Our calculated
temperature-dependent NPFs are bigger up to an order of magnitude as
compared to those by \cite{Rau00} for odd-A nuclei. However we note
that at higher temperatures, at times, the NPF calculated by
\cite{Rau03} is bigger by as much as an order of magnitude or more
due to neglect of high temperature effects in our calculation. We
further note that the calculated mass fractions depend heavily on
the number of nuclei considered in the ensemble (see
Eq.~(\ref{mass})). Whereas we took 89 nuclei in our smaller ensemble
to compare our results with those of \cite{Liu}, our calculated mass
fractions could have increased by 1-3 orders of magnitude or more by
choosing a much smaller ensemble (keeping other parameters of the
calculation constant). It is to be noted that size of ensemble and
its nuclear content were not mentioned by \cite{Liu}. Lastly we
attribute this difference to the microscopic calculation of energy
levels (till an excitation energy of 3 MeV) in our model. We also
incorporated measured energy levels in our calculation. For
even-even nuclei the two models evidently calculate much different
energy levels leading to the orders of magnitude differences. Our
calculation with new features is in somewhat better agreement with
that of \cite{Liu} for odd-A odd-Z nucleus $^{55}$Co. Within an
isotopic chain, it is further noted that the two mass fraction
calculations get in better agreement for heavier isotopes.

\section{Conclusions}
We presented here the mass fraction calculation in presupernova
cores as a function of lepton content of the stellar matter. We
assumed completion of silicon burning stage in our calculation which
led us to exploit the conditions of NSE and use of Saha equation for
the calculation of mass fractions. Our ensemble contained a total of
728 nuclei up to mass number 100. There are two distinguishing
features of our mass fraction calculation. Firstly we included
discrete energy levels (experimental and theoretical using the
pn-QRPA model) up to 10 MeV excitation energy scale in the nucleus
for the calculation of NPF. From 10-20 MeV we assumed a simple
nuclear level density function and used method of integration to
account for the continuum energy levels. Satisfactory convergence
was achieved in our calculation of the NPF for presupernova
conditions. Secondly we did not take the nucleus to be point-like
and rather assumed spherical Wigner-Seitz cells for every nucleus.
We then took into consideration the electron screening effect and
inserted the Coulomb correction term in the binding energy of the
nucleons. The resulting mass fractions were up to four orders of
magnitude smaller than those calculated by \cite{Liu} and may
significantly affect the rate of change of lepton-to-baryon ratio
(\.{Y$_{e}$}) of stellar matter during presupernova evolution of
massive stars. Fine-tuning of \.{Y$_{e}$} is one of the keys to
generate a successful explosion in core-collapse simulation. We are
in a process of calculating \.{Y$_{e}$} for typical presupernova
conditions and hope to report this as a future assignment.

Whereas we presented the mass fractions for only the most abundant
nuclei, the evolution of mass fractions with increasing $Y_{e}$, for
all 728 nuclei under presupernova conditions, may be requested as
ASCII files from the corresponding author. It is hoped that the mass
fraction calculation presented here would prove useful for the
analysis of stellar matter during presupernova phases of massive
stars and for related network calculations.

\acknowledgments J.-U. Nabi wishes to acknowledge the support
provided by the Higher Education Commission (Pakistan) through the
HEC Project No. 20-3099. The work of A. Tawfik and N. Ezzelarab was
supported by the World Laboratory for Cosmology and Particle Physics
(WLCAPP).

\clearpage \onecolumn
\newpage
\begin{figure}[h]
\begin{center}
\includegraphics[width=1.0\textwidth]{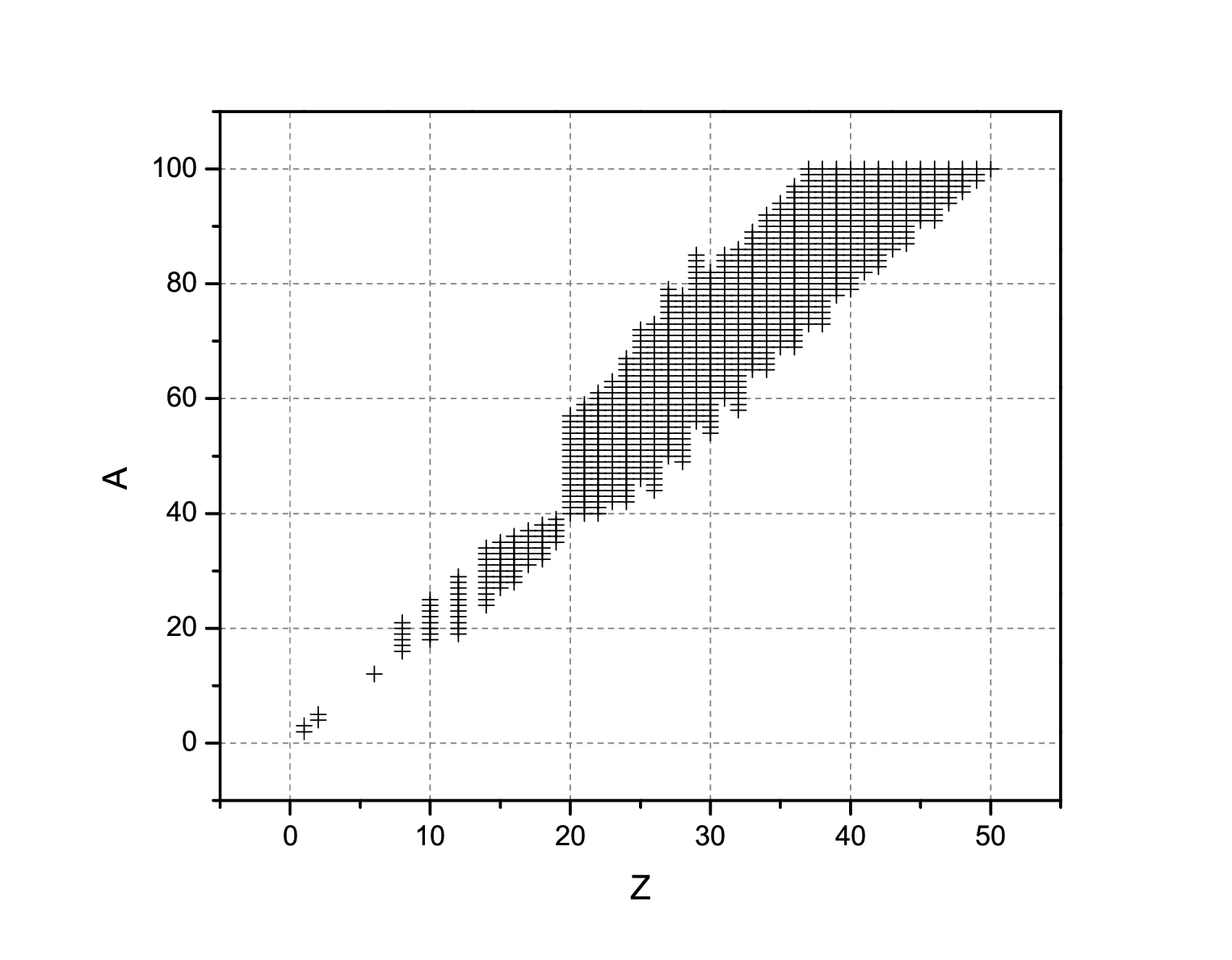}
\caption{ Nuclei present in our ensemble for mass fraction
calculation. }\label{fig0}
\end{center}
\end{figure}

\begin{figure}[h]
\begin{center}
\includegraphics[width=1.0\textwidth]{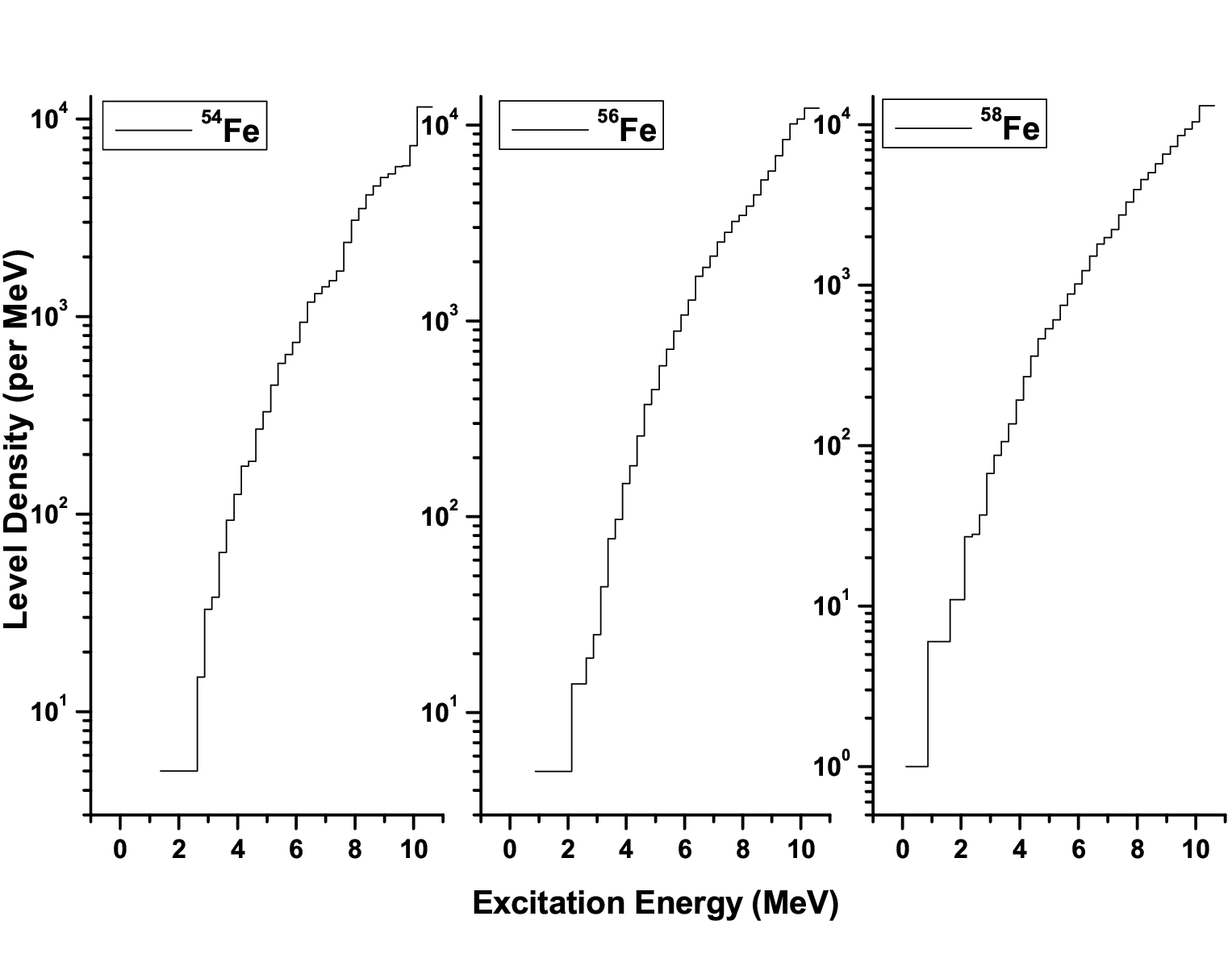}
\caption{ Calculated nuclear level densities for iron isotopes.
}\label{fig1}
\end{center}
\end{figure}

\begin{figure}[h]
\begin{center}
\includegraphics[width=1.0\textwidth]{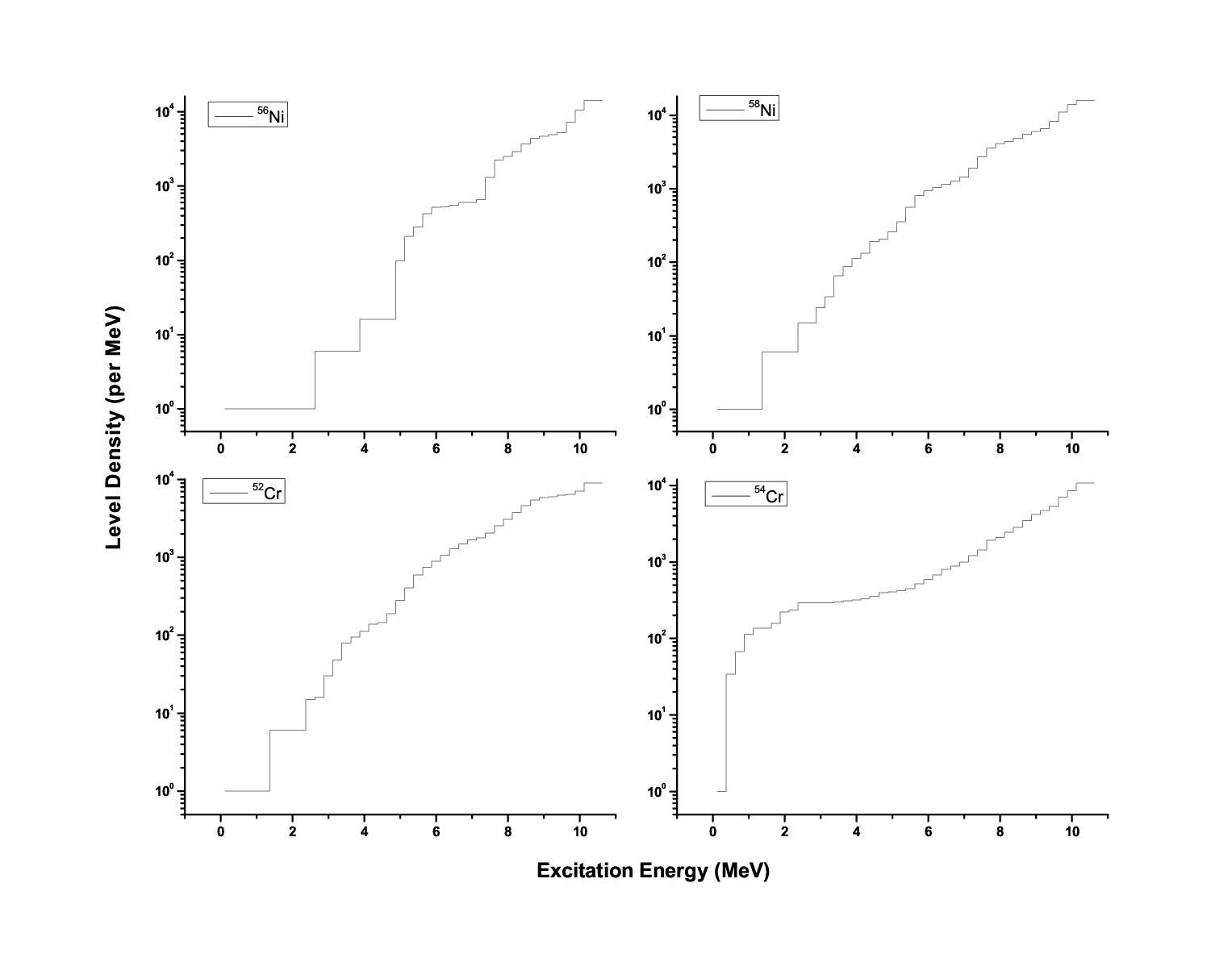}
\caption{ Calculated nuclear level densities for Fe-region nuclei.
}\label{fig2}
\end{center}
\end{figure}

\begin{figure}[h]
\begin{center}
\includegraphics[width=1.0\textwidth]{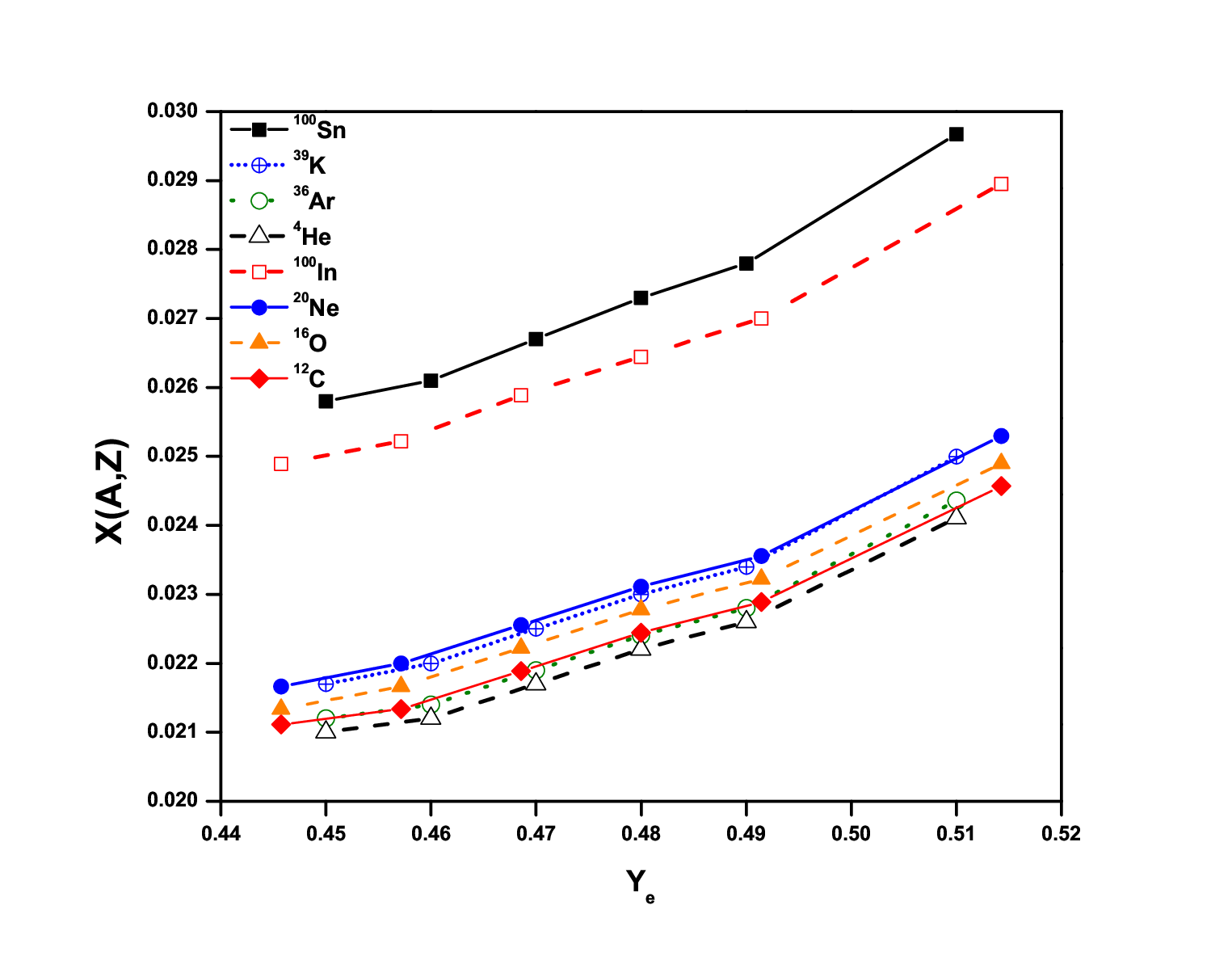}
\caption{ Mass fractions of the eight most abundant nuclei as
function of $Y_{e}$ at $T = 5\times10^{9}$K and $\rho_{st} =
1\times10^{7}$ g/cm$^{3}$ (color online). }\label{fig3}
\end{center}
\end{figure}

\begin{figure}[h]
\begin{center}
\includegraphics[width=1.0\textwidth]{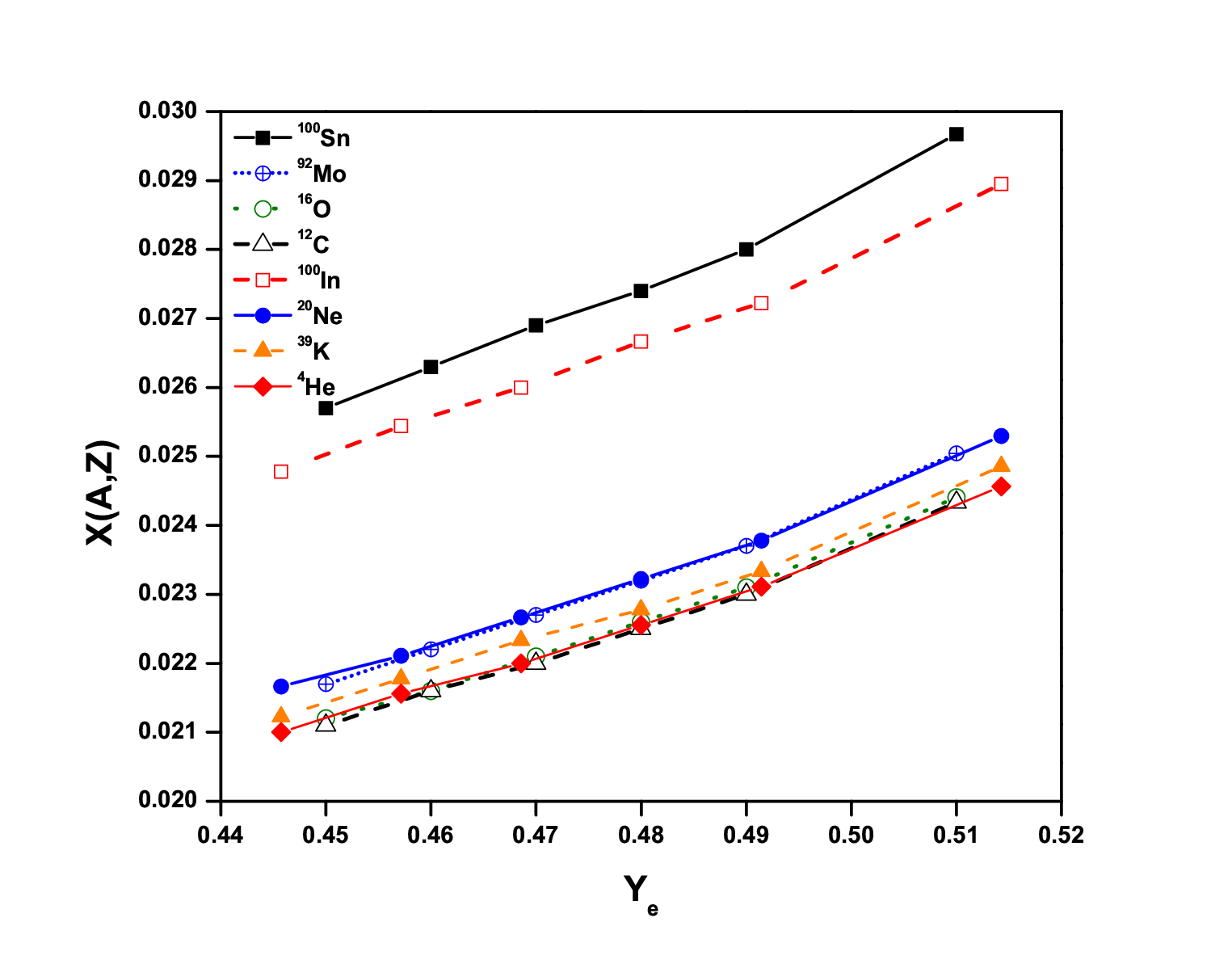}
\caption{ Mass fractions of the eight most abundant nuclei as
function of $Y_{e}$ at $T = 5\times10^{9}$K and $\rho_{st} =
1\times10^{9}$ g/cm$^{3}$ (color online). }\label{fig4}
\end{center}
\end{figure}

\begin{figure}[h]
\begin{center}
\includegraphics[width=1.0\textwidth]{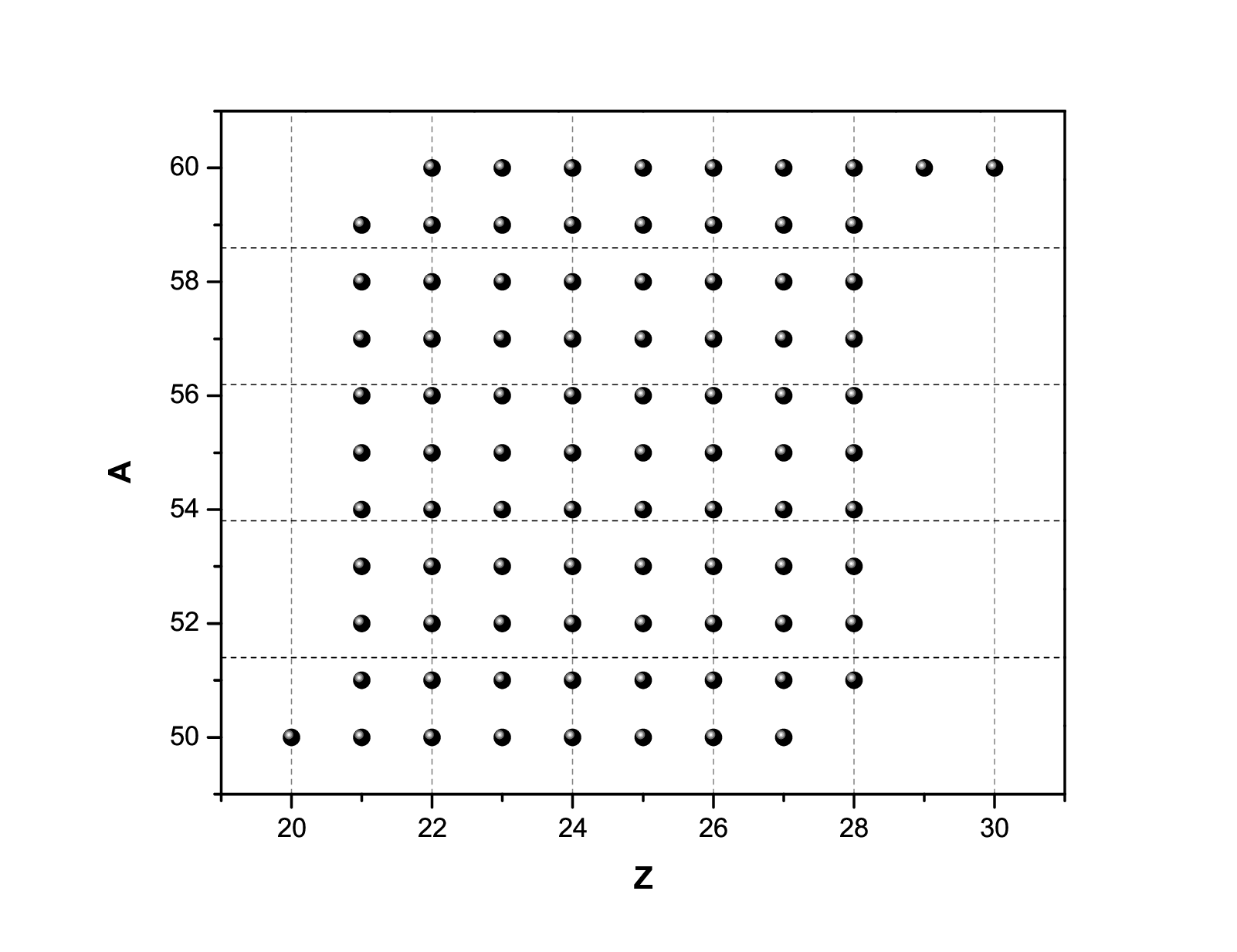}
\caption{ Nuclei present in our smaller ensemble for mass fraction
calculation with revised features. }\label{fig5}
\end{center}
\end{figure}


\begin{table}[h]
\scriptsize \caption{Calculated temperature dependent nuclear
partition functions for $^{50,51}$Sc, $^{50,51}$Ti, $^{50,51}$V,
$^{50,51}$Cr, $^{50,51}$Mn, $^{50,51}$Fe, $^{50,51}$Co and
$^{51,52}$Ni compared with the earlier statistical calculation of
\cite{Rau00} denoted by RT. Numbers in parenthesis indicate stellar
temperature in units of $10^{9}$K.} \label{ta1}
\begin{center}
\begin{tabular}{c|c|c|c|c|c|c}

    Nuclei& This work &RT        &This work  &RT         &This work  &RT  \\
          &G (1.0)    &G (1.0)    &G (3.0)      &G (3.0)       &G (10.0)     &G (10.0)\\
          &G (2.0)    &G (2.0)    &G (5.0)      &G (5.0)       &G (30.0)     &G (30.0)\\
    \hline
         &1.219E+01   &1.037E+00  &1.818E+01   &1.391E+00     &5.307E+01    &4.1E+00\\
$^{50}$Sc&1.499E+01   &1.207E+00  &2.452E+01   &1.723E+00 &1.456E+05
&....... \\
&   &  &   &     &    &\\&   &  &   & && \\
         &1.153E+01   &1.000E+00  &1.665E+01   &1.057E+00     &3.447E+01    &5.1E+00\\
$^{51}$Sc&1.470E+01   &1.008E+00  &1.999E+01   &1.373E+00     &3.170E+03    &....... \\
&   &  &   &     &    &\\&   &  &   & && \\
         &1.000E+00   &1.000E+00  &1.000E+00   &1.013E+00     &2.503E+00    &3.8E+00\\
$^{50}$Ti&1.000E+00   &1.001E+00  &1.006E+00   &1.165E+00     &3.365E+03     &.......\\
&   &  &   &     &    &\\&   &  &   & && \\
         &4.840E+00   &1.000E+00  &5.510E+00   &1.018E+00     &2.214E+01     &4.1E+00\\
$^{51}$Ti&5.113E+00   &1.001E+00  &6.898E+00   &1.197E+00     &3.224E+04     & ......\\
&   &  &   &     &    &\\&   &  &   & && \\
         &2.972E+00   &1.091E+00  &1.531E+01   &1.876E+00     &1.106E+02     &6.9E+00\\
$^{50}$V &8.773E+00   &1.461E+00  &3.139E+01   &2.745E+00     &1.345E+05     &.......\\
&   &  &   &     &    &\\&   &  &   & && \\
         &9.025E+00   &1.018E+00  &1.320E+01   &1.235E+00     &4.156E+01     &3.4E+00\\
$^{51}$V &1.092E+01   &1.120E+00  &1.830E+01   &1.488E+00     &3.108E+04     &.......\\
&   &  &   &     &    &\\&   &  &   & && \\
         &1.000E+00   &1.001E+00  &1.248E+00   &1.248E+00     &6.918E+00     &7.6E+00\\
$^{50}$Cr&1.053E+00   &1.053E+00  &1.956E+00   &1.964E+00     &3.139E+03     &.......\\
&   &  &   &     &    &\\&   &  &   & && \\
         &1.379E+00   &1.000E+00  &1.948E+00   &1.067E+00     &8.825E+00     &4.4E+00\\
$^{51}$Cr&1.692E+00   &1.011E+00  &2.434E+00   &1.381E+00     &2.918E+04     &.......\\
&   &  &   &     &    &\\&   &  &   & && \\
         &3.107E+00   &1.048E+00  &1.062E+01   &6.825E+00     &6.270E+01     &4.3E+01\\
$^{50}$Mn&6.671E+00   &4.233E+00  &2.002E+01   &1.223E+01     &1.215E+05     &.......\\
&   &  &   &     &    &\\&   &  &   & && \\
         &2.055E+00   &1.085E+00  &5.324E+00   &1.562E+00     &2.897E+01     &5.1E+00\\
$^{51}$Mn&3.806E+00   &1.339E+00  &8.501E+00   &2.015E+00     &2.999E+04     & ......\\
&   &  &   &     &    &\\&   &  &   & && \\
         &1.001E+00   &1.000E+00  &1.266E+00   &1.025E+00     &6.249E+00     &8.3E+00\\
$^{50}$Fe&1.059E+00   &1.002E+00  &1.989E+00   &1.336E+00     &3.125E+03     &.......\\
&   &  &   &     &    &\\&   &  &   & && \\
         &7.364E+00   &1.064E+00  &1.037E+01   &1.511E+00     &3.041E+01     &4.1E+00\\
$^{51}$Fe&8.863E+00   &1.296E+00  &1.383E+01   &1.890E+00     &2.993E+04     & ......\\
&   &  &   &     &    &\\&   &  &   & && \\
         &1.359E+01   &1.048E+00  &1.545E+01   &1.281E+00     &3.446E+01     &5.9E+00\\
$^{50}$Co&1.451E+01   &1.147E+00  &1.755E+01   &1.718E+00     &1.334E+05     & ......\\
&   &  &   &     &    &\\&   &  &   & && \\
         &8.233E+00   &1.000E+00  &8.831E+00   &1.020E+00     &1.558E+01     &2.4E+00\\
$^{51}$Co&8.556E+00   &1.003E+00  &9.317E+00   &1.117E+00     &2.909E+04     & ......\\
&   &  &   &     &    &\\&   &  &   & && \\
         &8.836E+00   &1.085E+00  &9.762E+00   &1.462E+00     &1.520E+01     &8.0E+00\\
$^{51}$Ni&9.405E+00   &1.250E+00  &1.026E+01   &2.095E+00     &3.107E+04     & .......\\
&   &  &   &     &    &\\&   &  &   & && \\
         &1.000E+00   &1.000E+00  &1.000E+00   &1.001E+00     &2.656E+00     &3.3E+00\\
$^{52}$Ni&1.000E+00   &1.000E+00  &1.013E+00   &1.032E+00     &3.671E+03     &......  \\
\hline
\end{tabular}
\end{center}
\end{table}

\begin{table}[h]
\caption{Calculated mass fractions for $^{52,53,54}$Fe, $^{55}$Co
and $^{56,58}$Ni at temperature $T= 5\times10^{9}$K, $\rho_{st} =
1\times10^{7}$ g.cm$^{-3}$ and $Y_{e}= 0.5$ compared with the
calculation of \cite{Liu}.} \label{ta2}
\begin{center}
\footnotesize\begin{tabular}{c|c|c|c} \hline
    Nuclei&$X(A,Z)$(this work with new features)&$X(A,Z)$(this work with revised features)&$X(A,Z)$\cite{Liu}\\
    \hline

$^{52}$Fe& 1.573$\times10^{-6}$& 5.901$\times10^{-6}$ &2.581$\times10^{-2}$\\
$^{53}$Fe&3.629$\times10^{-5}$&1.361$\times10^{-4}$  &3.653$\times10^{-2}$\\
$^{54}$Fe& 6.792$\times10^{-3}$&2.547$\times10^{-2}$  &2.984$\times10^{-1}$\\
$^{55}$Co&1.085$\times10^{-3}$& 2.689$\times10^{-2}$ &2.868$\times10^{-2}$\\
$^{56}$Ni& 4.545$\times10^{-5}$&1.705$\times10^{-4}$  &3.978$\times10^{-1}$\\
$^{58}$Ni&8.769$\times10^{-3}$& 3.288$\times10^{-2}$ &1.065$\times10^{-1}$\\

\hline
\end{tabular}
\end{center}
\end{table}

\begin{table}
\begin{center}
\caption {Comparison of calculated temperature-dependent  partition
functions $(G)$ with those of \cite{Rau00} for $^{52,53,54}$Fe,
$^{55}$Co and $^{56,57,58}$Ni. Numbers in parenthesis give stellar
temperature in units of 10$^{9}$ K.  } \tiny {
\begin{tabular}{c|c|c|c|c|c|c}
\hline
\textbf{Nuclei} & \textbf{G(0.01,0.5,1)}  & \textbf{G(0.01,0.5,1)} & \textbf{G(2,4,5)}  & \textbf{G(2,4,5)} & \textbf{G(6,8,10)}  & \textbf{G(6,8,10)} \\
 & QRPA & FRDM & QRPA  & FRDM & QRPA  & FRDM \\
 \hline
 $^{52}$Fe & 1.00E+00        & 1.00E+00 & 1.04E+00        & 1.04E+00 & 2.14E+00        & 2.10E+00 \\
           & 1.00E+00        & 1.00E+00 & 1.44E+00        & 1.44E+00 & 3.38E+00        & 3.00E+00 \\
           & 1.00E+00        & 1.00E+00 & 1.75E+00        & 1.74E+00 & 6.01E+00        & 4.57E+00 \\
    \hline
  $^{53}$Fe & 8.00E+00       & 1.00E+00 & 8.74E+00        & 1.01E+00 & 1.37E+01        & 1.46E+00 \\
            & 8.10E+00       & 1.00E+00 & 1.03E+01        & 1.14E+00 & 2.01E+01        & 2.13E+00 \\
            &8.33E+00        & 1.00E+00 & 1.16E+01        & 1.27E+00 & 3.24E+01        & 3.58E+00\\
     \hline
 $^{54}$Fe & 1.00E+00        & 1.00E+00 & 1.00E+00        & 1.00E+00 & 1.60E+00        & 1.55E+00 \\
           & 1.00E+00        & 1.00E+00 & 1.10E+00        & 1.10E+00 & 3.18E+00        & 2.75E+00 \\
           & 1.00E+00        & 1.00E+00 & 1.27E+00        & 1.26E+00 & 7.04E+00        & 5.54E+00 \\
     \hline
 $^{55}$Co & 8.00E+00        & 1.00E+00 & 8.73E+00        & 1.00E+00 & 1.03E+01        & 1.05E+00 \\
          & 8.05E+00         & 1.00E+00 & 9.43E+00        & 1.00E+00 & 1.30E+01        & 1.26E+00 \\
          & 8.26E+00         & 1.00E+00 & 9.78E+00        & 1.01E+00 & 2.12E+01        & 1.96E+00 \\
          \hline
 $^{56}$Ni& 1.00E+00         & 1.00E+00 & 1.00E+00        & 1.00E+00 & 1.05E+00        & 1.03E+00 \\
          & 1.00E+00         & 1.00E+00 & 1.00E+00        & 1.00E+00 & 1.41E+00        & 1.18E+00 \\
          & 1.00E+00         & 1.00E+00 & 1.01E+00        & 1.01E+00 & 2.97E+00        & 1.67E+00 \\
          \hline
 $^{58}$Ni& 1.00E+00         & 1.00E+00 & 1.00E+00        & 1.00E+00 & 1.57E+00        & 1.51E+00 \\
          & 1.00E+00         & 1.00E+00 & 1.09E+00        & 1.09E+00 & 3.14E+00        & 2.71E+00 \\
          & 1.00E+00         & 1.00E+00 & 1.25E+00        & 1.24E+00 & 7.28E+00        & 5.75E+00 \\
\end{tabular}}
\label{ta3}
\end{center}
\end{table}

\begin{table}
\caption {Comparison of calculated temperature-dependent  partition
functions $(G)$ with those of \cite{Rau03} for $^{52,53,54}$Fe,
$^{55}$Co and $^{56,57,58}$Ni. Numbers in parenthesis give stellar
temperature in units of 10$^{9}$ K.    } \tiny {
\begin{tabular}{c|c|c|c|c|c|c|c|c|c}
\hline \textbf{Nuclei}    & \textbf{G(12,14,16)} &
\textbf{G(12,14,16)} &\textbf{G(12,14,16)} & \textbf{G(18,20,22)} &
\textbf{G(18,20,22)} & \textbf{G(18,20,22)} &\textbf{G(24,26,28)} & \textbf{G(24,26,28)} & \textbf{G(24,26,28)} \\
   & QRPA & ETFSI &FRDM & QRPA & ETFSI & FRDM & QRPA & ETFSI & FRDM \\
\hline
$^{52}$Fe & 1.19E+01 & 7.72E+00 & 8.23E+00 & 1.13E+02 & 1.11E+02 & 1.39E+02 & 8.03E+02 & 2.29E+03 & 3.19E+03\\
          & 2.49E+01 & 1.64E+01 & 1.86E+01 & 2.30E+02 & 3.05E+02 & 3.99E+02 & 1.37E+03 & 6.15E+03 & 8.84E+03\\
          & 5.32E+01 & 4.09E+01 & 4.91E+01 & 4.43E+02 & 8.40E+02 & 1.14E+03 & 2.21E+03 & 1.63E+04 & 2.42E+04\\
          \hline
 $^{53}$Fe& 5.87E+01 & 7.07E+00 & 6.88E+00 & 7.21E+02 & 8.28E+01 & 8.22E+01 & 6.80E+03 & 1.23E+03 & 1.31E+03\\
          & 1.24E+02 & 1.52E+01 & 1.48E+01 & 1.67E+03 & 2.01E+02 & 2.04E+02 & 1.22E+04 & 3.06E+03 & 3.37E+03\\
          & 2.97E+02 & 3.48E+01 & 3.40E+01 & 3.51E+03 & 4.96E+02 & 5.14E+02 & 2.03E+04 & 7.63E+03 & 8.72E+03\\
          \hline
$^{54}$Fe & 1.54E+01 & 1.25E+01 & 1.23E+01 & 1.36E+02 & 2.11E+02 & 2.09E+02 & 9.07E+02 & 4.23E+03 & 4.47E+03\\
          & 3.24E+01 & 3.04E+01 & 2.97E+01 & 2.69E+02 & 5.72E+02 & 5.76E+02 & 1.53E+03 & 1.14E+04 & 1.24E+04\\
          & 6.71E+01 & 7.86E+01 & 7.70E+01 & 5.08E+02 & 1.56E+03 & 1.60E+03 & 2.46E+03 & 3.08E+04 & 3.46E+04\\
          \hline
$^{55}$Co& 4.48E+01 & 4.02E+00 & 3.86E+00 & 7.66E+02 & 5.62E+01 & 5.45E+01 & 7.49E+03 & 9.39E+02 & 9.80E+02 \\
          & 1.13E+02 & 9.23E+00 & 8.81E+00 & 1.80E+03 & 1.42E+02 & 1.41E+02 & 1.34E+04 & 2.44E+03 & 2.63E+03 \\
          & 2.99E+02 & 2.25E+01 & 2.15E+01 & 3.85E+03 & 3.64E+02 & 3.69E+02 & 2.24E+04 & 6.35E+03 & 7.11E+03 \\
          \hline
$^{56}$Ni& 7.81E+00 & 3.20E+00 & 3.23E+00 & 1.19E+02 & 6.76E+01 & 7.17E+01 & 9.74E+02 & 1.74E+03 & 2.01E+03 \\
          & 2.06E+01 & 8.03E+00 & 8.19E+00 & 2.59E+02 & 2.01E+02 & 2.19E+02 & 1.70E+03 & 5.10E+03 & 6.08E+03\\
          & 5.12E+01 & 2.28E+01 & 2.37E+01 & 5.21E+02 & 5.94E+02 & 6.64E+02 & 2.79E+03 & 1.48E+04 & 1.83E+04\\
          \hline
$^{58}$Ni& 1.67E+01 & 1.38E+01 & 1.38E+01 & 1.64E+02 & 2.98E+02 & 3.10E+02 & 1.15E+03 & 7.47E+03 & 8.42E+03\\
          & 3.69E+01 & 3.66E+01 & 3.69E+01 & 3.30E+02 & 8.72E+02 & 9.29E+02 & 1.96E+03 & 2.18E+04 & 2.54E+04\\
          & 7.89E+01 & 1.03E+02 & 1.05E+02 & 6.33E+02 & 2.55E+03 & 2.80E+03 & 3.16E+03 & 6.35E+04 & 7.67E+04\\
          \hline
 \end{tabular}}
\label{ta4}
\end{table}

\end{document}